\documentclass[a4paper,11pt]{article}
\pdfoutput=1 

\usepackage{jheppub} 

\usepackage[T1]{fontenc} 

\usepackage{epsfig}
\usepackage{graphicx,color}

\newcommand{\beq}{\begin{equation}}
\newcommand{\eeq}{\end{equation}}
\newcommand{\be}{\begin{eqnarray}}
\newcommand{\ee}{\end{eqnarray}}

\title{
Constraints on the dark Z model
    from the Higgs boson phenomenology
}

\author[a,b]{Dong-Won Jung,}
\author[c]{Kang Young Lee}
\author[c,1]{and Chaehyun Yu \note{Corresponding author.}}

\affiliation[a]{Department of Physics, Yonsei University, Seoul 03722, Korea}
\affiliation[b]{Department of Physics, Chonnam National University, 
Gwangju 61186, Korea}
\affiliation[c]{ Department of Physics Education \& RINS, 
Gyeongsang National University, Jinju 52828, Korea}

\emailAdd{dongwon.jung@yonsei.ac.kr}
\emailAdd{kylee.phys@gnu.ac.kr}
\emailAdd{chyu@kias.re.kr}

\abstract{
We study constraints on the hidden sector model mediated 
by an additional SU(2) Higgs doublet
from the phenomenology of Higgs bosons.
The hidden sector is assumed to contain a hidden U(1) gauge symmetry
and the hidden U(1) gauge boson gets the mass
by the electroweak symmetry breaking to be a dark $Z$ boson.
The Higgs sector of the model is similar to that of 
the two Higgs doublet model of type I
except for the absence of the CP-odd scalar boson.
Using the programs of {\tt HiggsBounds} and {\tt HiggsSignals},
we incorporate current experimental limits from LEP, Tevatron and LHC
to examine the Higgs sector in our model
and derive constraints on model parameters.
We also discuss the implications of the model 
on the dark matter phenomenology.
}
\begin{document}
\maketitle

\section{Introduction}

Following the discovery of the Higgs boson 
\cite{Higgs-atlas,Higgs-cms},
extensive study on the scalar sector 
has been conducted to reveal
the detailed properties of the Higgs boson
at the Large Hadron Collider (LHC).
Although the Higgs mechanism with a SU(2)$_L$ doublet Higgs field
provides a tenable explanation of 
the electroweak symmetry breaking (EWSB)
in the Standard Model (SM),
still exist many theoretical drawbacks such as naturalness problem,
flavour structure, and origin of neutrino masses
related to the Higgs sector.
Therefore many new physics models beyond the SM
suggest extensions of the Higgs sector.
So far no significant deviations from the SM 
are shown in experiments
and the precise measurements of Higgs boson sectors
provide strong constraints 
on the theoretical structure of new physics (NP).

It is another shortcoming of the SM
that the SM does not include 
the dark matter (DM) candidates of the Universe,
which are demanded for explanation of the present
astrophysical and cosmological observations.
Various kinds of DM candidates
have emerged from theoretical models beyond the SM
\cite{Bertone:2004pz}.
A class of DM candidates is predicted 
as natural ingredients of the models,
while other class of DM candidates is assumed
to be included in the `hidden' sector
which is very weakly interacting with the SM.

Recently we have suggested a hidden sector model
where a SM singlet fermion in the hidden sector is a DM candidate \cite{Jung:2020ukk}.
In this model, an additional Higgs doublet is 
introduced to connect the hidden sector to the SM sector
and a hidden U(1)$_X$ gauge symmetry is also introduced
for the new Higgs doublet to couple to the singlet fermion.
The hidden sector fermion as well as the additional Higgs doublet
have the U(1)$_X$ charge but the SM fields do not.
The U(1)$_X$ gauge symmetry is spontaneously broken
by the EWSB in this model.
Thus there exists an extra massive gauge boson $Z'$
mixed with the SM $Z$ boson as a result.
We call it a dark $Z$ boson,
since its couplings to the SM fermions are 
proportional to those of the $Z$ boson 
with the suppression factor of the mixing angle.
The dark $Z$ boson mass is of order the EW scale,
and could be much light in this model.
Hence, the dark $Z$ boson affects the electroweak processes
and is stringently constrained 
by the collider and the low energy neutral current (NC) experiments
\cite{Jung:2020ukk,Jung:2021bjt}.

Since we introduce one more Higgs doublet
as the mediator field between the hidden sector
and the SM sector,
we have two SU(2) scalar doublets in the model and
the Higgs structure is similar to that of the two Higgs doublet model (2HDM).
Due to the U(1)$_X$ charge, the new Higgs doublet
cannot couple to the SM fermions
and consequently the Higgs sector of our model is same
as that of the type-I 2HDM. The discrete symmetry in the typical 2HDM 
is absent in the model and the Yukawa structure is
controlled by the $U(1)_X$ gauge symmetry as in Ref.~\cite{Ko:2013zsa}.

After the EWSB,
there exist three additional Higgs bosons, 
one CP-even neutral Higgs boson
and a pair of charged Higgs bosons.
Note that no CP-odd neutral Higgs boson appear in this model
since corresponding degree of freedom is accountable
for the dark $Z$ boson mass.
In this paper we carry out an intensive study on the model
with the phenomenology of Higgs bosons 
and the electroweak processes.
The model parameters in the Higgs sector
are two masses of neutral and charged Higgs bosons,
$m_h$ and $m_\pm$,
and two mixing angles 
$\alpha$ and $\beta$ for them, respectively.
We probe consistency of these parameters with
large number of experimental data
using the public codes 
{\tt HiggsBounds} \cite{Bechtle:2020pkv}
and {\tt HiggsSignals} \cite{Bechtle:2020uwn}.
The former provides the 95\% exclusion limit
for the extra scalar production 
in collider experiments while the latter provides
$\chi^2$ and P-values of a model with respect to
observables of the SM-like Higgs boson at the LHC.

The dark $Z$ boson, which is dubbed as $Z'$, 
is a portal to the hidden sector where DM is living. 
We assume the DM candidate is a Dirac fermion.
The U(1)$_X$ charge of the DM fermion 
is independent of the visible sector. 
The interaction strength of the DM fermion to the $Z'$
and the DM mass 
are new free parameters 
to adjust the observed relic density 
and the discovery limit for the DM-nucleon cross sections.
With a comprehensive numerical analysis, 
we show that significant portion of the parameter space 
can explain the known DM properties 
in relation with cosmology and astrophysics.
We also provide arguments about the validity of the results considering 
the astrophysical phenomena such as supernova cooling.

This paper is organized as follows.
The model is explained in section \ref{model}.
In section \ref{analyses} we analyze this model mainly focusing
on the Higgs sector. We describe the analysis of the phenomenology of the Higgs bosons 
with {\tt HiggsBounds} and {\tt HiggsSignals} in section \ref{higgspheno}.
Discussion about constraints on the model parameters 
together with the electroweak processes,
the atomic parity violation, the $\rho$ parameter,
and the $Z$ boson total width is followed in section \ref{rho}, \ref{APV} and \ref{Zdecay}, respectively.
We show that our dark matter candidate is acceptable
for the relic density of the universe
and the present direct detection experiments
in section \ref{DMpheno}.
Section \ref{conclusion} is devoted to the summary of the results and conclusion.

\section{The Model \label{model}}

We assume that the hidden sector consists of 
a Dirac fermion with U(1)$_X$ gauge symmetry.
The hidden sector fermion is a SM gauge singlet and
the dark matter candidate, 
of which gauge charge is assigned to be
$ \psi_X (1, 1, 0, X)$
with the U(1)$_X$ charge $X$.
The SM fields do not have the U(1)$_X$ charge
and do not couple to the hidden sector directly.
Thus we require a mediator field between
the SM and the hidden sector.
We introduce an extra Higgs doublet $H_1$ as the mediator
and as a result we have two Higgs doublets in this model. 
The gauge charges of the mediator $H_1$ and 
the SM Higgs doublet $H_2$ are assigned to be
\be
H_1 \left(1, 2, \frac{1}{2}, \frac{1}{2} \right),~~~
H_2 \left(1, 2, \frac{1}{2}, 0 \right),
\ee
under the gauge group 
${\rm SU}(3)_c \times {\rm SU}(2)_L \times {\rm U}(1)_Y 
\times {\rm U}(1)_X$.
The kinetic mixing of U(1)$_X$ with the SM U(1)$_Y$ is ignored here.

We write the lagrangian of the Higgs sector as
\be
{\cal L}_H = (D^\mu H_1)^\dagger D_\mu H_1 
	   + (D^\mu H_2)^\dagger D_\mu H_2 - V(H_1, H_2) 
	   + {\cal L}_{\rm Y}(H_2), 
\ee
where $V(H_1, H_2)$ is the Higgs potential and
${\cal L}_{\rm Y}$ the Yukawa interactions,
\be
-{\cal L}_Y = g_{ij}^d \bar{Q}^i_L H_2 d^j_R
              + g_{ij}^u \bar{Q}^i_L \tilde{H}_2 u^j_R
              + g_{ij}^l \bar{L}^i_L H_2 l^j_R   + H.C. 
\ee
with $\tilde{H}_2 = i \sigma_2 H^*$.
The covariant derivative is written by
\be
D^\mu &=& \partial^\mu + i g W^{\mu\;\!a} T^a
                 + i g' B^\mu Y + i g_X A_X^\mu X.
\ee
Due to the U(1)$_X$ charge,
$H_1$ does not couple to the SM fermions
and the flavour structure of this model 
is naturally that of the 2HDM of type I.
The Higgs potential is given by
\be
V(H_1,H_2) &=& \mu_1^2 H_1^\dagger H_1 + \mu_2^2 H_2^\dagger H_2
\nonumber \\
	   &+& \lambda_1 (H_1^\dagger H_1)^2
	      + \lambda_2 (H_2^\dagger H_2)^2
              + \lambda_3 (H_1^\dagger H_1)(H_2^\dagger H_2)
              + \lambda_4 (H_1^\dagger H_2)(H_2^\dagger H_1),
\ee
where the soft breaking term and the $\lambda_5$ terms are 
prevented by the U(1)$_X$ charge.

After the EWSB, two vacuum expectation values (VEVs) are evolved,
\beq
H_i = \left( \begin{array}{c} H_i^+ \\[1pt]
          \frac{1}{\sqrt{2}} (v_i+\rho_i+i \eta_i)
              \end{array} \right),
\eeq
where $v_1^2 + v_2^2 = v^2$.
We define the ratio $\tan \beta \equiv v_2/v_1$.
The U(1)$_X$ gauge boson $A_X$ as well as the SM gauge bosons
get masses from the EWSB.
Keeping the photon $A$ as a massless mode, 
we find that
the extra massive neutral gauge boson $Z'$ is mixed 
with the ordinary $Z$ boson to yield the physical states 
\be
\left( \begin{array}{c}
        A_X \\[1pt] W^3 \\[1pt] B \end{array} \right)
= \left( \begin{array}{c}
              c_X Z' + s_X Z \\[1pt]
             -s_X c_W Z' + c_X c_W Z + s_W A \\[1pt]
              s_X s_W Z' - c_X s_W Z + c_W A \end{array} \right),
\ee
where $s_X = \sin \theta_X$, $c_X = \cos \theta_X$
with the $Z$-$Z'$ mixing angle $\theta_X$,
and $s_W = \sin \theta_W$, $c_W = \cos \theta_W$
with the Weinberg angle $\theta_W$.
The gauge boson masses are given by
\be
m_{Z,Z'}^2 = \frac{1}{8} \left(
          g_X^2 v_1^2 + (g^2 + {g'}^2) v^2 \pm
          \sqrt{(g_X^2 v_1^2 - (g^2 + {g'}^2) v^2)^2
               + 4 g_X^2 (g^2 + {g'}^2) v_1^4} \right),
\ee
and the mixing angle $\theta_X$ by
\beq
\tan 2 \theta_X = \frac{-2 g_X \sqrt{g^2 + {g'}^2} v^2 \cos^2 \beta}
                       {(g^2+{g'}^2)v^2 - g_X^2 v^2 \cos^2 \beta}.
\eeq
Two new parameters, the U(1)$_X$ gauge coupling $g_X$ and 
$\tan \beta$ are introduced in the electroweak sector.
We take two observables, the dark $Z$ mass $m_{Z'}$ and
$\tan \beta$ as free parameters
by translating $g_X$ into $m_{Z'}$.
Then the mixing angle can be expressed by
\be
s_X^2 = \frac{m_W^2 \cos^2 \beta}
         {c_W^2 (m_Z^2-m_{Z'}^2)-m_W^2 \cos^2 \beta}
        \frac{m_{Z'}^2}{m_Z^2-m_{Z'}^2}
\ee
in terms of observables.

We write the NC interactions
in terms of the physical gauge bosons as
\be
{\cal L}_{NC} 
= - e A^\mu \bar{f} Q \gamma_\mu f
    - c_X Z^\mu \left( g_V \bar{f} \gamma_\mu f
          + g_A \bar{f} \gamma_\mu \gamma_5 f \right)
    + s_X {Z'}^\mu \left( g_V \bar{f} \gamma_\mu f
          + g_A \bar{f} \gamma_\mu \gamma_5 f \right),
\nonumber \\
\ee
where the couplings $e$, $g_V$, $g_A$ are
defined in the same manner as those of the SM.
Note that the $Z'$ couplings are same as the ordinary $Z$ couplings
up to the suppression factor $\tan \theta_X $.
Therefore we call the extra gauge boson $Z'$ the dark $Z$ boson.

After the EWSB, we diagonalize the mass matrices 
of the neutral and charged Higgs bosons 
to derive physical states and to get the physical masses.
We have two CP-even neutral scalar bosons
of which masses are given by
\be
m_{h,H}^2 = \lambda_1 v_1^2 + \lambda_2 v_2^2 \pm
           \sqrt{ (\lambda_1 v_1^2 - \lambda_2 v_2^2)^2
                  +(\lambda_3 + \lambda_4)^2 v_1^2 v_2^2},
\label{neutralmasses}%
\ee
and the mixing angle $\alpha$ given by
\be
\tan 2\alpha = \frac{-(\lambda_3 + \lambda_4) \tan \beta}
                    {\lambda_1 - \lambda_2 \tan^2 \beta}.
\ee
The physical states of the neutral Higgs bosons are defined by
\be
        \left( \begin{array}{c}
        \rho_1  \\[1pt]
        \rho_2  \\[1pt] \end{array}
        \right)
      = \left( \begin{array}{cc}
        \cos \alpha &\  \sin \alpha \\[1pt]
       -\sin \alpha &\  \cos \alpha \\[1pt] \end{array}
        \right)
        \left( \begin{array}{c}
        h  \\[1pt]
        H  \\[1pt] \end{array}
        \right)
      = \left( \begin{array}{c}
        h \cos \alpha + H \sin \alpha  \\[1pt]
       -h \sin \alpha + H \cos \alpha  \\[1pt] \end{array}
        \right).
\ee
Note that $H$ is the SM-like Higgs
and $h$ the extra neutral Higgs boson.

Diagonalizing the charged states, 
we have a pair of charged Higgs bosons, $H^\pm$, of which masses are
\be
m_\pm^2 = -\frac{1}{2} \lambda_4 v^2,
\label{chargedmass}%
\ee
and a pair of massless modes,
where the mixing angle is
\be
\tan 2 \beta = \frac{2 M_{12}^2}{M_{11}^2 - M_{22}^2 }
              = \frac{2 v_1 v_2}{v_1^2-v_2^2}.
\ee
The charged states are given
in terms of $H_1^\pm,H_2^\pm$,
\be
        \left( \begin{array}{c}
        H_1^\pm  \\[1pt]
        H_2^\pm  \\[1pt] \end{array}
        \right)
      = \left( \begin{array}{cc}
        \cos \beta &\ -\sin \beta \\[1pt]
        \sin \beta &\  \cos \beta \\[1pt] \end{array}
        \right)
        \left( \begin{array}{c}
        G^\pm  \\[1pt]
        H^\pm  \\[1pt] \end{array}
        \right)
      = \left( \begin{array}{c}
        G^\pm \cos \beta - H^\pm \sin \beta  \\[1pt]
        G^\pm \sin \beta + H^\pm \cos \beta  \\[1pt] \end{array}
        \right),
\ee
where the massless modes $G^\pm$ are corresponding to
the longitudinal modes of the $W^\pm$ bosons.

\section{Analyses \label{analyses}}

In this section, we analyze our model by considering both theoretical constraints
and experimental limits from various experiments. 

\subsection{Higgs phenomenology \label{higgspheno}}

There are two CP-even neutral scalar bosons and
a pair of charged scalar bosons after the EWSB in this model.
One neutral scalar is the SM-like Higgs
of which mass is fixed as the measured value,
$ m_H = 125.10$ GeV.
The other neutral scalar boson may be either lighter
or heavier than the SM-like Higgs boson
and its mass $m_h$ is a free parameter of the model.
The other free parameters of the Higgs sector 
are the charged Higgs mass $m_\pm$, 
and the mixing angles $\alpha$ and $\beta$.

In order to analyze the Higgs potential,
we write the Higgs quartic couplings 
in terms of masses and mixing angles,
\be
\lambda_1 &=& \frac{1}{2v_1^2}
       \left( m_h^2 \cos^2 \alpha + m_H^2 \sin^2 \alpha \right),
\nonumber \\
\lambda_2 &=& \frac{1}{2v_2^2}
       \left( m_h^2 \sin^2 \alpha + m_H^2 \cos^2 \alpha \right),
\nonumber \\
\lambda_3 &=& \frac{1}{v^2}
       \left( 2 m_\pm^2
       + \frac{\sin 2 \alpha}{\sin^2 \beta}
           \left( m_H^2 - m_h^2 \right) \right),
\nonumber \\
\lambda_4 &=& -\frac{2 m_{\pm}^2}{v^2} .
\ee
The quartic couplings are assumed to be 
in the perturbative domain,
$|\lambda_i|<4 \pi$ in this analysis.
The potential should be bounded from below
with the following conditions 
\be
\lambda_1 > 0,~~~~~
\lambda_2 > 0,~~~~~
\lambda_3 > -2\sqrt{\lambda_1 \lambda_2},~~~~~
\lambda_3 + \lambda_4 > -2\sqrt{\lambda_1 \lambda_2},
\ee
for the stable vacuum. 
We also require the perturbative unitarity of the $WW$ scattering
at tree level, which is translated into the following conditions \cite{Lee:1977eg,Muhlleitner:2016mzt}
\be
&&|\lambda_1| < 4 \pi,~~~~~
|\lambda_2| < 4 \pi,~~~~~
|\lambda_3| < 8 \pi,
\nonumber \\
&&| \lambda_3 + \lambda_4 | < 8 \pi,~~~~~
| \lambda_3 + 2 \lambda_4 | < 8 \pi,
\nonumber \\
&&|\lambda_1 + \lambda_2 
\pm\sqrt{(\lambda_1-\lambda_2)^2+\lambda_4^2}| < 8 \pi, 
\nonumber \\
&&|3(\lambda_1 + \lambda_2) 
\pm\sqrt{9(\lambda_1-\lambda_2)^2+(2\lambda_3+\lambda_4)^2}| < 8 \pi, 
\nonumber \\
&&\sqrt{\frac{2\lambda_3^2+4\lambda_3 \lambda_4+\lambda_4^2}{2}} < 8 \pi. 
\ee

Lots of data from Higgs search experiments
and the measurements of the SM-like Higgs boson
has been provided by collider experiments, 
at LEP, Tevatron and LHC.
We implement the analysis 
to achieve allowed parameter sets of the extended Higgs sector 
by applying experimental limits
using the public codes 
{\tt HiggsBounds} \cite{Bechtle:2020pkv}
which compares various predictions in the Higgs sector
with experimental data.
{\tt HiggsBounds} includes the exclusion limits
at the 95$\%$ C.L. for the extended Higgs boson searches.
We also use {\tt HiggsSignals} \cite{Bechtle:2020uwn}
which perform a $\chi^2$ test of the Higgs sector predictions
against the signal rate and mass measurements from colliders.
The code {\tt HiggsSignals} computes the signal strengths with model parameters.

For {\tt HiggsBounds} and {\tt HiggsSignals},
we express the effective couplings $\kappa_{ijk}$
in terms of following independent model parameters
\be 
(m_{Z'}, \sin \alpha, \tan \beta, m_{h}, m_\pm ),
\label{eq:parameters}%
\ee
where $m_h$ and $m_\pm$ are defined in Eq.~(\ref{neutralmasses}) and (\ref{chargedmass}), respectively.
The effective couplings are normalized 
by the corresponding SM couplings.

The scalar to fermion pair couplings are given by
\be
\kappa_{H\bar{f}f} = \frac{\cos \alpha}{\sin \beta},~~~~~
\kappa_{h\bar{f}f} = -\frac{\sin \alpha}{\sin \beta},
\ee
where the SM couplings are $g_{H\bar{f}f}=-m_f/v$. 
The triple couplings of one scalar to two gauge bosons
normalized by the SM Higgs couplings
$g_{HWW}^{\rm SM}=g m_W$ and $g_{HZZ}^{\rm SM}=g m_Z/c_W$, 
respectively, are read as
\be
\kappa_{HW^+ W^-} &=& \sin (\alpha+\beta),
\nonumber  \\
\kappa_{hW^+ W^-} &=& \cos (\alpha+\beta),
\ee
and
\be
\kappa_{HZZ} &=& c_X^2 \sin (\alpha+\beta)
     - \cos \beta \sin \alpha \left(
                \frac{2 g_X}{g} c_W c_X s_X 
	       - \left( \frac{g_X}{g} \right)^2 c_W^2 s_X^2 \right),
\nonumber  \\
\kappa_{hZZ} &=& c_X^2 \cos (\alpha+\beta)
     - \cos \beta \cos \alpha \left(
                \frac{2 g_X}{g} c_W c_X s_X 
	       - \left( \frac{g_X}{g} \right)^2 c_W^2 s_X^2 \right),
\nonumber  \\
\kappa_{HZZ'} &=& -c_X s_X \sin (\alpha+\beta)
     - \cos \beta \sin \alpha \left(
                \frac{g_X}{g} c_X^2 
                - \frac{g_X}{g} c_W s_X^2 
	       - \left( \frac{g_X}{g} \right)^2 c_W^2 c_X s_X \right),
\nonumber  \\
\kappa_{hZZ'} &=& -c_X s_X \cos (\alpha+\beta)
     - \cos \beta \cos \alpha \left(
                \frac{g_X}{g} c_X^2 
                - \frac{g_X}{g} c_W s_X^2 
	       - \left( \frac{g_X}{g} \right)^2 c_W^2 c_X s_X \right),
\nonumber  \\
\kappa_{HZ' Z'} &=& s_X^2 \sin (\alpha+\beta)
     + \cos \beta \sin \alpha \left(
                \frac{2 g_X}{g} c_W c_X s_X 
	       + \left( \frac{g_X}{g} \right)^2 c_W^2 c_X^2 \right),
\nonumber  \\
\kappa_{hZ' Z'} &=& s_X^2 \cos (\alpha+\beta)
     + \cos \beta \cos \alpha \left(
                \frac{2 g_X}{g} c_W c_X s_X 
	       + \left( \frac{g_X}{g} \right)^2 c_W^2 c_X^2 \right).
\ee
We find that
both the SM-like Higgs boson and the additional neutral scalar boson 
can decay into $Z Z^\prime$ and $Z^\prime Z^\prime$ modes
depending on $m_{Z'}$ and $m_h$
which do not exist in the typical 2HDM.
These channels are open due to 
the U(1)$_X$ charge assignment for the $H_1$ 
as well as the mixing between the $Z$ and $Z^\prime$ bosons. 
Thus these decay channels must be taken into account 
for the calculation of the branching ratios of the scalar bosons 
which are input information 
for the {\tt HiggsBounds} and {\tt HiggsSignals}. Furthermore,
these channels can probe distinct signals of this model from the typical 2HDMs with discrete $Z_2$ symmetry as well as the SM.

The triple gauge couplings read as
\be
\kappa_{W^+ W^- Z} = c_X, ~~~~~~~
\kappa_{W^+ W^- Z'} = -s_X, 
\ee
and the Higgs triple couplings as
\be
\kappa_{HHH} &=& 3 \left(
        2 \lambda_1 \sin^3 \alpha \cos \beta
        + 2 \lambda_2 \cos^3 \alpha \sin \beta
	+ (\lambda_3+\lambda_4) \cos \alpha \sin \alpha \cos (\alpha-\beta)
	\right),
\nonumber  \\
\kappa_{hhh} &=& 3 \left(
        2 \lambda_1 \cos^3 \alpha \cos \beta
        - 2 \lambda_2 \sin^3 \alpha \sin \beta
	+ (\lambda_3+\lambda_4) \cos \alpha \sin \alpha \cos (\alpha-\beta)
	\right),
\nonumber  \\
\kappa_{Hhh} &=&  
            6 \cos \alpha \sin \alpha 
	      (\lambda_1 \cos \alpha \cos \beta
	         + \lambda_2 \sin \alpha \sin \beta) \nonumber \\
            &&+ (\lambda_3+\lambda_4)
	      (\sin (\alpha+\beta) 
	      -3 \cos \alpha \sin \alpha \cos (\alpha-\beta)) ,
\nonumber  \\
\kappa_{HHh} &=&  
            6 \cos \alpha \sin \alpha 
	      (\lambda_1 \sin \alpha \cos \beta
	         - \lambda_2 \cos \alpha \sin \beta) \nonumber \\
            &&+ (\lambda_3+\lambda_4)
	      (\cos (\alpha+\beta) 
	      -3 \cos \alpha \sin \alpha \sin (\alpha-\beta)), 
\ee
with normalization of $v^2/m_Z^2$.
Note that the SM coupling is $g_{HHH}^{\rm SM} = 3 m_H^2/v=6 \lambda v$.

Since there are no charged Higgs bosons in the SM,
the effective couplings with the charged Higgs are defined by
appropriate normalizations.
The charged Higgs to fermions couplings are given by
\be
\kappa_{H^\pm \bar{f}f} &=& \frac{1}{\tan \beta},
\ee
with the normalization $\sqrt{2} m_f/v$.
We write the couplings of the charged Higgs bosons to neutral gauge bosons as
\be
g_{Z H^\pm H^\mp} &=& \frac{1}{c_W} \left( (1-2s_W^2) c_X 
              + \frac{g_X}{g} c_W s_X \sin^2 \beta \right) , ~~~
\nonumber  \\
g_{Z' H^\pm H^\mp} &=& \frac{1}{c_W} \left( -(1-2s_W^2) s_X 
              + \frac{g_X}{g} c_W c_X \sin^2 \beta \right) 
\ee
without normalization.
Note that $g_{Z H^\pm H^\mp} = (1-2s_W^2)/c_W$ in the typical 2HDM.
The neutral and charged Higgs couplings 
associated by the $W^\pm$ boson are given by
\be
\kappa_{H H^\pm W^\mp} &=&\cos(\alpha+\beta),~~~
\nonumber \\
\kappa_{h H^\pm W^\mp} &=&-\sin(\alpha+\beta),
\ee
and the neutral and charged gauge boson couplings
associated with the $H^\pm$ given by
\be
\kappa_{Z H^\pm W^\mp} &=& -\left( \frac{g_X}{g} \right) 
                             \cos \beta \sin \beta c_W s_X, ~~~
\nonumber \\
\kappa_{Z' H^\pm W^\mp} &=& -\left( \frac{g_X}{g} \right) 
                             \cos \beta \sin \beta c_W c_X. 
\ee
The Higgs triple couplings with the charged Higgs are
\be
\kappa_{HH^+ H^-} &=& 2 \lambda_1 \sin \alpha \cos \beta \sin^2 \beta
                  + 2 \lambda_2 \cos \alpha \sin \beta \cos^2 \beta
      + \lambda_3 (\cos \alpha \sin^3 \beta+\sin \alpha \cos^3 \beta)
\nonumber \\
&&~~~~~~~  - \lambda_4 (\cos \alpha \cos^2 \beta \sin \beta
                             +\sin \alpha \cos \beta \sin^2 \beta),
\nonumber \\
\kappa_{hH^+ H^-} &=& 2 \lambda_1 \cos \alpha \cos \beta \sin^2 \beta
                  - 2 \lambda_2 \sin \alpha \sin \beta \cos^2 \beta
      + \lambda_3 (\cos \alpha \cos^3 \beta-\sin \alpha \sin^3 \beta)
\nonumber \\
&&~~~~~~~  + \lambda_4 (\sin \alpha \cos^2 \beta \sin \beta
                             -\cos \alpha \cos \beta \sin^2 \beta)
\ee
with normalization of $v^2/m_Z^2$.

We require that randomly generated free parameters of 
(\ref{eq:parameters})
have to pass the exclusion limit of {\tt HiggsBounds} and 
then the P-value evaluated from {\tt HiggsSignals} is less than $0.05$.

\subsection{$\Delta \rho$ constraints \label{rho}}

Since our model contains the dark $Z$ boson,
even possibly light, 
the electroweak phenomena would be affected in general.
First, we consider the constraints arising 
from the electroweak precision observable, $T$.
The $Z$ boson mass is shifted 
from the definition of the SM in this model, 
\be
m_Z^2 &=& \frac{m_W^2}{c_W^2 c_X^2} - m_{Z'}^2 \frac{s_X^2}{c_X^2},
\ee
at tree level, while the $W^\pm$ mass remains unchanged. 
The $\rho$ parameter is defined 
by the ratio of $W$ and $Z$ boson masses,
$ \rho \equiv m_W^2/m_Z^2 c_W^2$, 
which is expected to be 1 in the SM.
The correction $\Delta \rho_X$ parametrizes the NP effects
\be
\Delta \rho_X \equiv 1 - \frac{1}{\rho}
     \approx -s_X^2 \left(1 - \frac{m_{Z'}^2 c_W^2}{m_W^2} \right),
\ee
in the leading order of $s_X$.
There also exist the new scalar contributions, $\Delta \rho_H$, 
to the $\rho$ parameter at loop levels.
We calculate $\Delta \rho_H$ in our model such that \cite{Grimus:2007if}
\be
\Delta \rho_H &=& \frac{\alpha}{16 \pi^2 m_W^2 s_W^2}
   \left( \sin^2 (\alpha+\beta) F(m_\pm^2, m_H^2)
         + \cos^2 (\alpha+\beta) F(m_\pm^2, m_h^2)
	 \right.
\nonumber \\
&&~~~~	 + 3 \cos^2 (\alpha+\beta) F(m_{Z'}^2, m_H^2)
	 + 3 \sin^2 (\alpha+\beta) F(m_{Z'}^2, m_h^2)
\nonumber \\
&&~~~~  \left. - 3 \cos^2 (\alpha+\beta) F(m_{W}^2, m_H^2)
	 - 3 \sin^2 (\alpha+\beta) F(m_{W}^2, m_h^2)
	 \right),
\ee
where 
\be
F(x,y) = \frac{x+y}{2} - \frac{xy}{x-y} \ln \frac{x}{y}.
\ee

The value of $\Delta \rho$ is obtained from 
the $T$ variable by the relation
\beq
\Delta \rho = \alpha(m_Z) T.
\eeq
The $T$ value 
\be
T = 0.09 \pm 0.07,
\ee
and the fine structure constant 
$\alpha^{(5)}(m_Z)^{-1} = 127.955 \pm 0.010$
are obtained from PDG \cite{PDG}, 
where the superscript $(5)$ denotes the five-loop result in QED.

\subsection{The atomic parity violation \label{APV}}

Since the left-handed and the right-handed couplings
of the $Z f \bar{f}$ vertex are different,
the $Z$ boson exchange produces
the parity violation of the atomic spectra.
Thus the measurement of the atomic parity violation (APV) 
constrains the NP effects in the NC interaction.
The effective lagrangian for the APV
is written as
\be
{\cal L} = -\frac{G_F}{\sqrt{2}} \left(
      g_{AV}^u ({\bar e} \gamma_\mu \gamma^5 e)({\bar u} \gamma^\mu u)
     + g_{AV}^d ({\bar e} \gamma_\mu \gamma^5 e)({\bar d} \gamma^\mu d)
	      \right),
\ee
where the nucleon couplings
$g_{AV}^p \equiv 2 g_{AV}^u+g_{AV}^d$ 
and $g_{AV}^n \equiv g_{AV}^u+2 g_{AV}^d$. 

We define the weak charge of the nuclei 
\beq
Q_W \equiv -2 \left( Z g_{AV}^p + N g_{AV}^n \right),
\eeq
where $Z$ ($N$) is the number of protons (neutrons) in the atom.
The weak charge $Q_W^{SM} \approx -N + Z (1-4 s_W^2)$ at tree level
in the SM.
Including the dark $Z$ contribution, we have
\beq
Q_W = Q_W^{SM} \left( 1+\frac{m_Z^2}{m_{Z'}^2} s_X^2 \right),
\eeq
in the leading order of $s_X$.
Using the SM value 
$Q_W^{SM} =  -73.16\pm0.03$ \cite{APVSM1,APVSM2},
the present experimental value for the Cs atom
is given by
\cite{APV}
\beq
Q_W^{exp} = -72.82\pm0.42,
\eeq
which yields the constraint on the model
\beq
\frac{m_Z^2}{m_{Z'}^2} s_X^2 \le 0.006,
\eeq
at 95 \% CL.

\subsection{$Z$ boson decays \label{Zdecay}}

Generically light are
the dark $Z$ boson and the additional neutral Higgs $h$
in this model.
When the sum of the masses of the dark $Z$ and $h$
is less than the $Z$ boson mass,
the $Z \to Z' h$ channel opens
and it contributes to the $Z$ boson total width.
The width of $Z \to Z' h$ decay is given by
\be
\Gamma(Z \to Z' h) &=& \frac{g^2 \kappa_{hZZ'}^2}
                            {192 \pi m_Z c_W^2 m_{Z'}^2}
               \left( m_h^4 - 2 m_h^2 m_Z^2 -2 m_h^2 m_{Z'}^2 +m_Z^4
                      +10 m_Z^2 m_{Z'}^2 +m_{Z'}^4 \right)
\nonumber \\
             && \times \left( 1-\frac{2 m_h^2}{m_Z^2}
                        -\frac{2 m_{Z'}^2}{m_Z^2}+\frac{m_h^4}{m_Z^4}
                        +\frac{m_{Z'}^4}{m_Z^4}-\frac{2 m_h^2 m_{Z'}^2}{m_Z^4}
                        \right).
\ee
The dark $Z$ decays are suppressed 
by the $Z$-$Z'$ mixing $\sin \theta_X$
and the $h$ decays suppressed by $\sin \alpha$.
Then the $Z'$ and $h$ live long and decay outside the detector when the DM particle mass is larger than half the $Z'$ mass. If the DM particle is light enough, the $Z'$ will decay into a pair of the DM particles. Thus the $Z \to Z' h$ decay will eventually contribute to the invisible $Z$ decay width in most cases.

The $Z$ total width $\Gamma_Z$ is precisely measured at LEP and SLC
\cite{ALEPH:2005ab}
\be
\Gamma_Z = 2.4955 \pm 0.0023 ~~{\rm GeV},
\ee
and shows a good agreement with the SM prediction,
$\Gamma_Z^{\rm SM} = 2.4941 \pm 0.0009$ GeV.
We add the partial width of $Z \to Z' h$ decay to the SM total width,
$\Gamma_Z^{\rm New} = \Gamma_Z^{\rm SM} + \Gamma(Z \to Z' h)$,
and constrain the model prediction by the experimental limits.
Since $\Gamma_Z$ is very precisely measured, 
the additional contribution $\Gamma(Z \to Z' h)$
should be highly suppressed.

\subsection{Results \label{results}}

\begin{figure}[t]
\centering
\includegraphics[width=12cm]{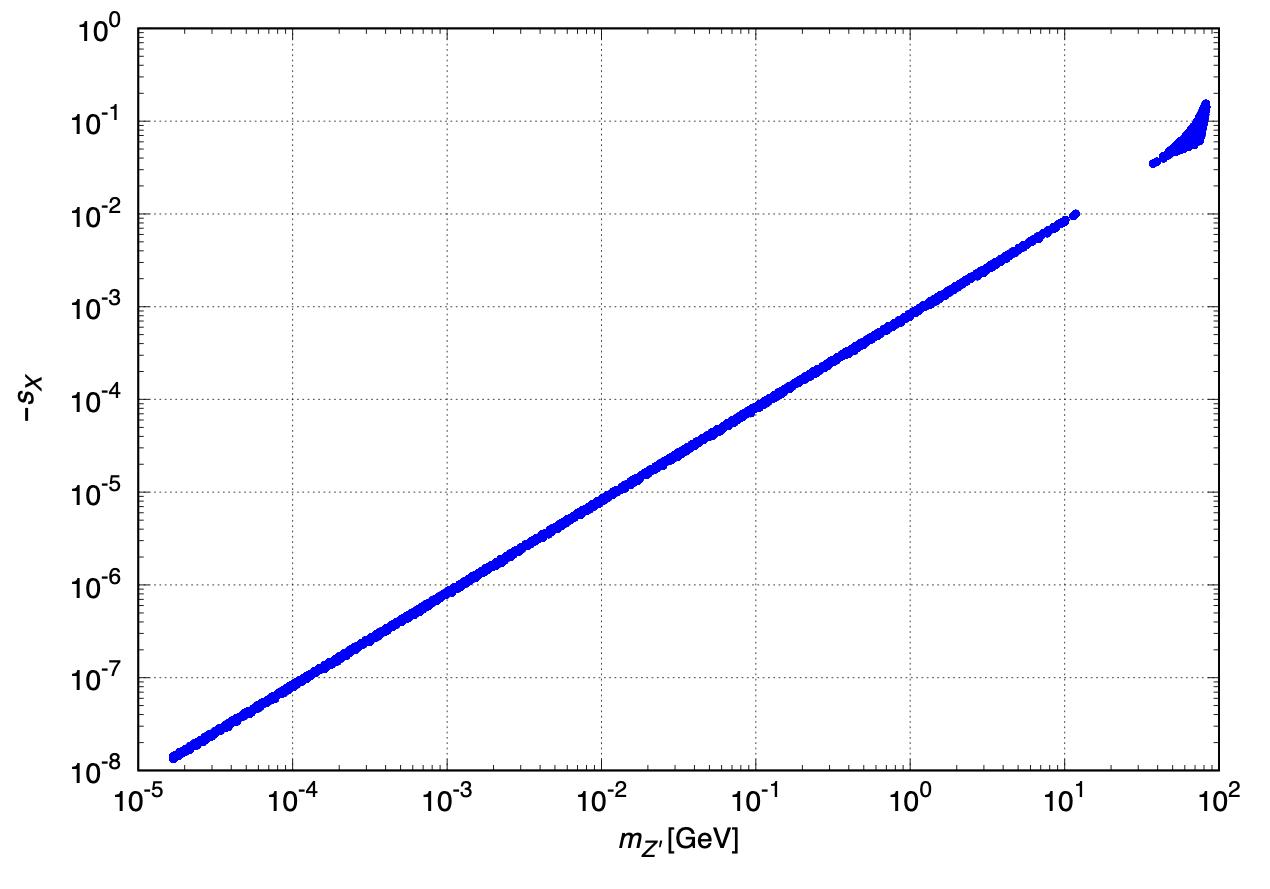}
\caption{
Allowed parameter set of $(m_{Z'}, |\sin \theta_X|)$.
}\label{fig:sX_ZX}
\end{figure}

\begin{figure}[t]
\centering
\includegraphics[width=7cm]{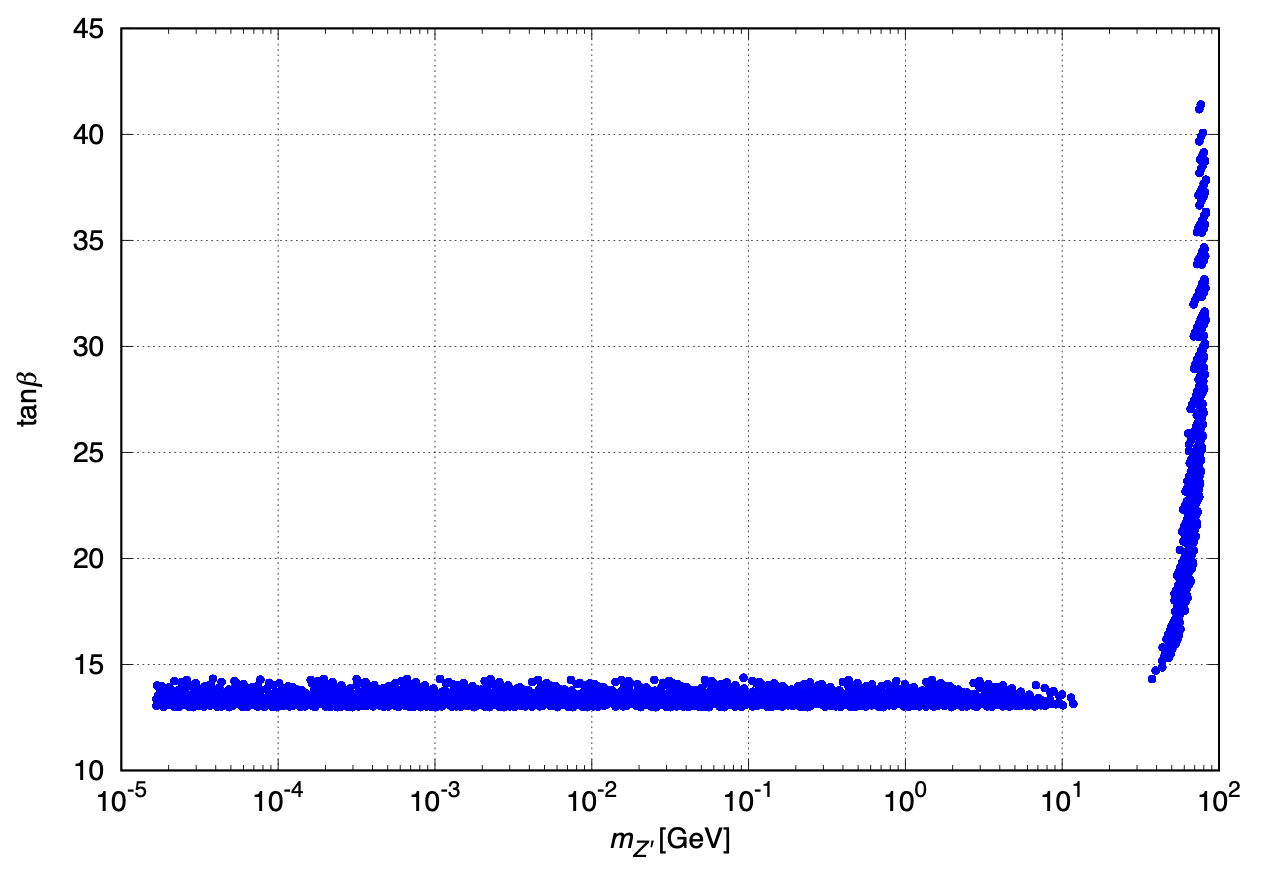}
\includegraphics[width=7cm]{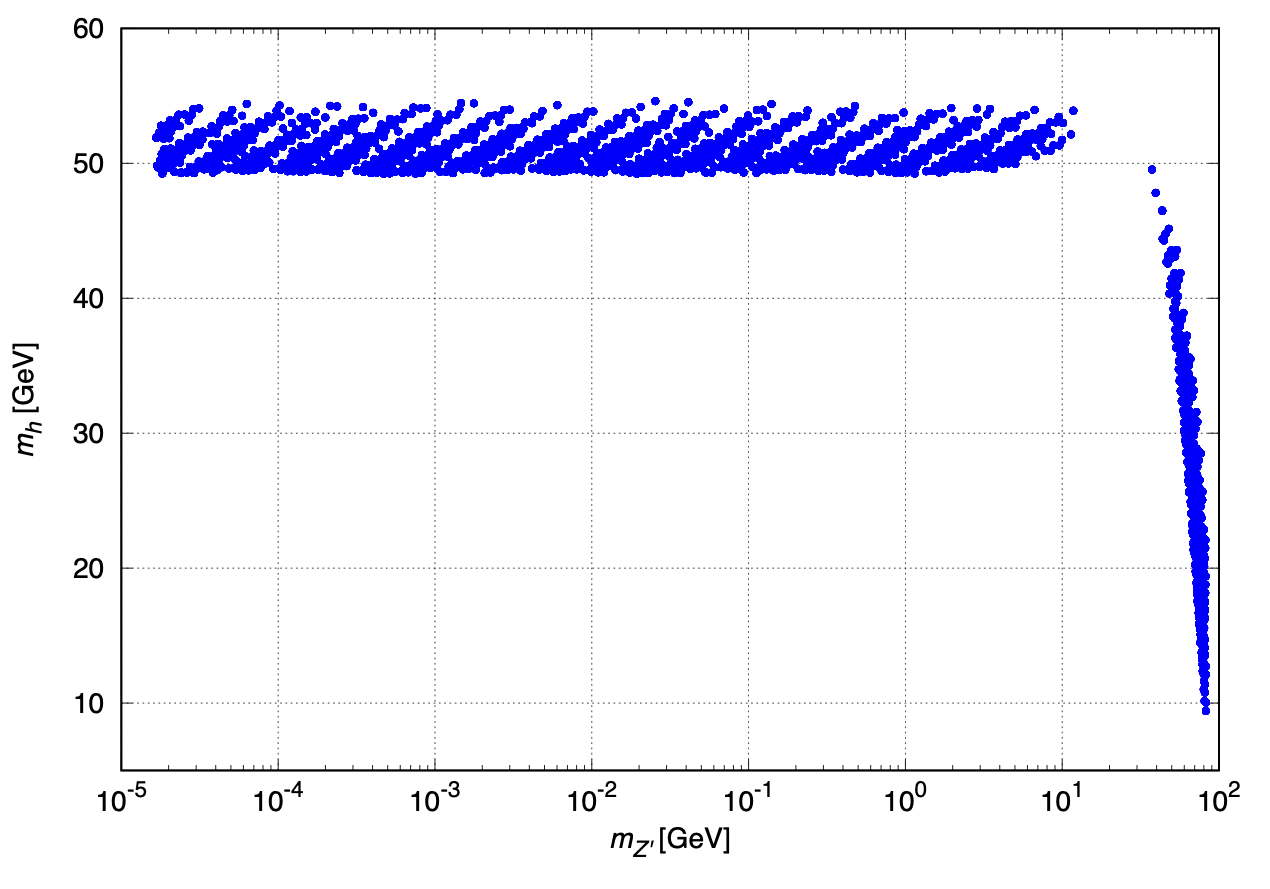}
\caption{
Allowed values of $\tan \beta$ (left) and $m_h$ (right) 
with respect to $m_{Z'}$.
}\label{fig:tbmh_Zp}
\end{figure}

\begin{figure}[t]
\centering
\includegraphics[width=12cm]{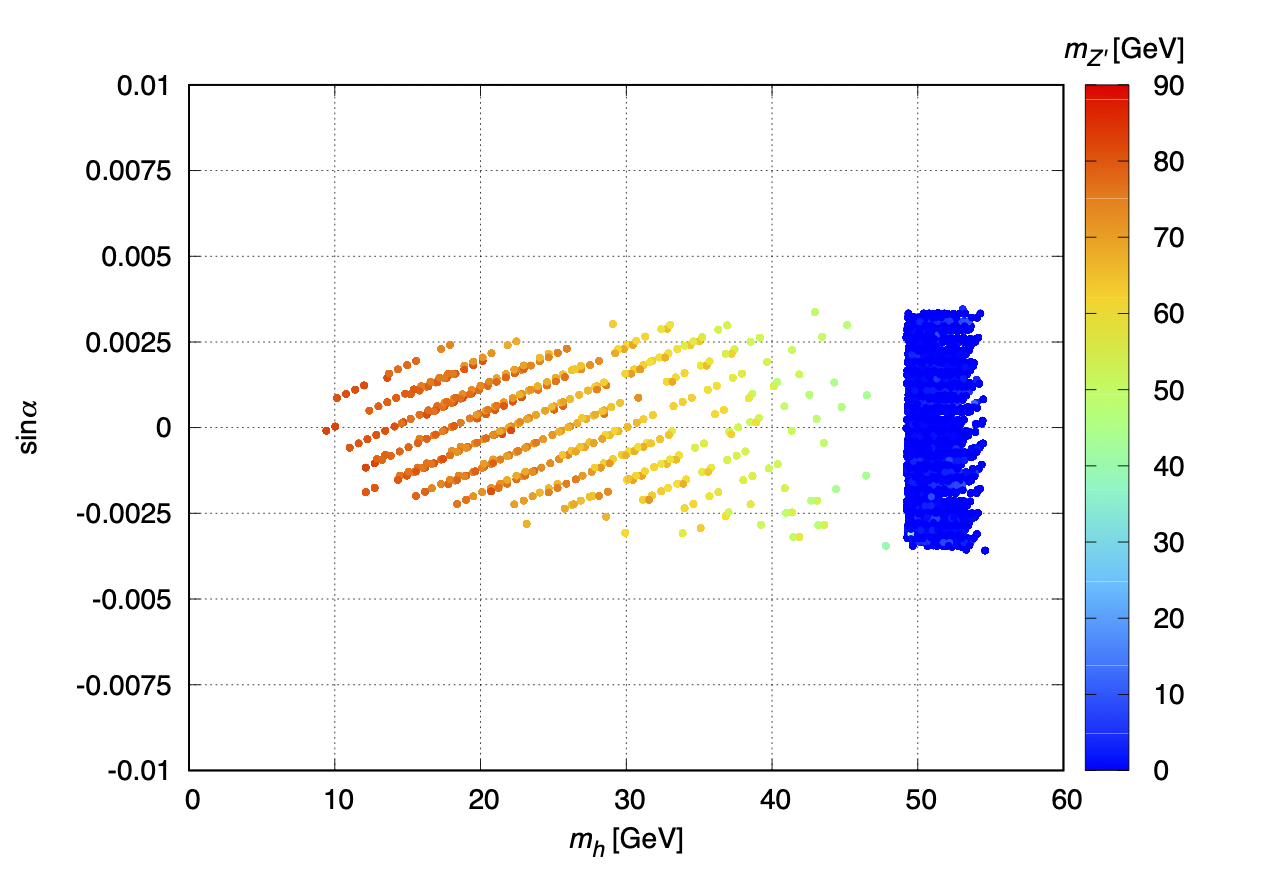}
\caption{
Allowed parameter set of $(m_{h}, \sin \alpha)$.
}\label{fig:mhsina}
\end{figure}

\begin{figure}[t]
\centering
\includegraphics[width=12cm]{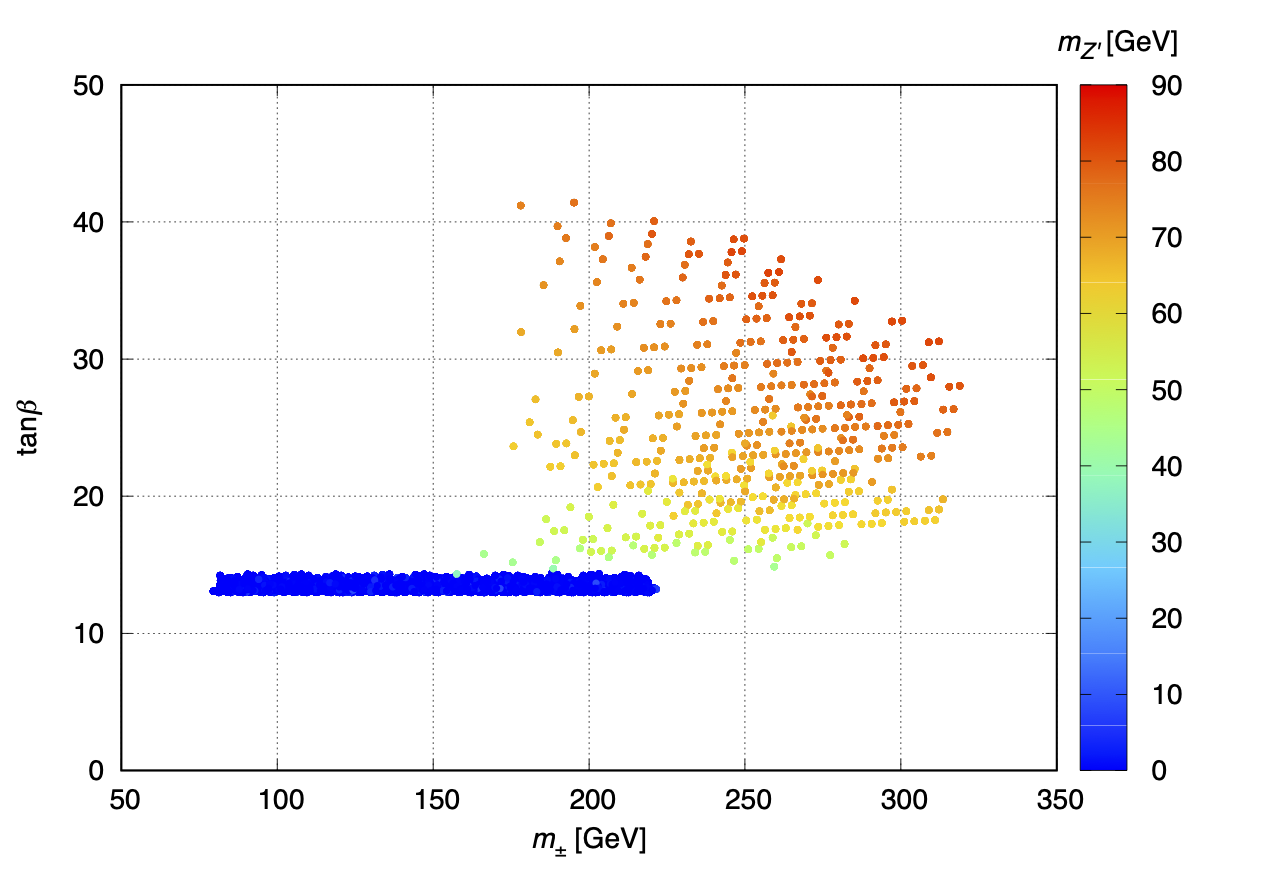}
\caption{
Allowed parameter set of $(m_{\pm}, \tan \beta)$.
}\label{fig:mchtanb}
\end{figure}

Figure~\ref{fig:sX_ZX} depicts the parameter set $(m_{Z'},|\sin \theta_X|)$
allowed by the {\tt HiggsBounds}, 
{\tt HiggsSignals}, and the electroweak constraints.
We see that there are two regions in the parameter space,
the light dark $Z$ region, $m_{Z'}<12$ GeV and
the heavy region, $m_{Z'}>30$ GeV.
The reason why there are two separate allowed parameter regions
is that each region corresponds to 
different suppression conditions of the $Z \to Z' h$ decay.
The light dark $Z$ region corresponds to 
the small mixing angle $|s_X|$ and consequently
the suppressed coupling of $\kappa_{hZZ'}$,
while the heavy region to
the kinematic suppression.
The kinematic suppression implies either
$m_Z \approx m_{Z'}+m_h$ or $m_Z < m_{Z'}+m_h$.
If $m_Z < m_{Z'}+m_h$, the decay channel does not open
and no constraints arise from the $Z$ total width.
When $m_Z \approx m_{Z'}+m_h$, although the decay channel opens,
$\Gamma(Z \to Z' h)$ is suppressed due to
the small kinematic factor of the decay width 
which is factorized by $(m_Z-m_{Z'}-m_h)$.

The values of $\tan \beta$ and $m_h$ 
are crucially constrained in the light dark $Z$ region
as is seen in Fig.~\ref{fig:tbmh_Zp}.
We find that $13 < \tan \beta < 15$
and 50 GeV $< m_h <$ 55 GeV
when $m_{Z'} < 12$ GeV.
On the other hand,
in the case of kinematic suppression, $m_Z \approx m_{Z'}+m_h$ or $m_Z < m_{Z'}+m_h$, wide range of parameters is allowed.
We see that $13 < \tan \beta < 42$
and 10 GeV $< m_h <$ 55 GeV.

The allowed mass and mixing angle 
of the additional neutral Higgs boson are shown in Fig.~\ref{fig:mhsina}.
The allowed mass of $h$ is 
$10 ~{\rm GeV} < m_h < 55$ GeV as also shown in Fig.~\ref{fig:tbmh_Zp}.
We see that the small mixing angle is preferred,
$|\sin \alpha |< 0.004$,
which implies that the fermion-coupled scalar $\rho_2$
is mostly the SM-like Higgs boson $H$.
Consequently the VEVs $v_2 \sim v$,
$\tan \beta$ is expected to be large
and the additional scalar is generically 
lighter than the SM-like Higgs.

We show the allowed charged Higgs mass and mixing angle in Fig.~\ref{fig:mchtanb}. 
Since $\tan \beta$ is large enough,
the model is free from
constraints from the $b \to s \gamma$ decays in $B$ physics
and $H^\pm \to \tau \nu$ and $H^\pm \to tb$ decays at the LHC.
We can see that there are also two groups of allowed points in Fig. 3 and Fig. 4, which are corresponding to the coupling suppression 
and the kinematic suppression for $Z \to Z' h$ decay, respectively, as explained above. 

Before moving to the DM phenomenology, we emphasize that our model is still viable as one of the simplest extension of the SM, even without the hidden sector. We should mention that 
the $Z$ boson decay width in this section does not include the channels of the $Z$ boson decay into particles in the hidden sector. Depending on the configuration of the hidden sector,
the $Z$ boson decay width could further constrain the parameter space.

\section{Dark Matter Phenomenology \label{DMpheno}}

We introduce a vector-like Dirac fermion with U(1)$_X$ gauge symmetry in the hidden sector, which is a candidate for DM. The hidden sector lagrangian is given by
\be
{\cal L}_{\rm hs} = -\frac{1}{4} F_X^{\mu \nu} F_{X \mu \nu} 
                    + \bar{\psi}_X i \gamma^\mu D_\mu \psi_X
                          - m_X \bar{\psi}_X \psi_X,
\ee
where
\be
D_\mu = \partial_\mu + i g_X A_\mu^X X,
\ee
and $X$ is the U(1)$_X$ charge and $m_X$ the mass for the $\psi_X$.
As a result of the spontaneous symmetry breaking and consequential $Z$-$Z'$ mixing, 
the DM interaction terms with the physical gauge bosons are reformulated as
\be
{\cal L}_{\rm DM}^{int} = i g_X\,X \,\bar{\psi}_X \gamma^\mu \psi_X
			  \left( c_X {Z'}_\mu + s_X Z_\mu \right).
\ee
Note that we have only two additional parameters, $m_X$ 
and $X$, which are independent of the structure of 
the visible sector.
Based on the Dirac fermionic DM scenario, we perform the phenomenological 
analysis on properties of DM with the \texttt{micrOMEGAs}
\cite{micromegas}, especially for the relic abundance and the DM-nucleon cross section for the allowed values of parameters in the previous section. Thermal freeze-out scenario is assumed for the relic 
abundance calculation.

The most recent DM contribution to the relic abundance density $\Omega$ 
is obtained from measurements of the anisotropy of the cosmic microwave background (CMB) and of the spatial distribution of galaxies
\cite{Planck:2018vyg}. The present value of the relic density for 
nonbaryonic DM is
\beq
\Omega_{\rm CDM} h^2 = 0.120 \pm 0.001.
\eeq
Such a precise value provides a stringent constraint
on the model parameters. We search for the parameter sets of the DM mass and 
its charge $X$ that make the relic abundance stay within 3-$\sigma$ range from 
the central value of the observation, with the perturbativity constraint $g_X X \le \sqrt{4\pi}$.
The acceptable relic density can be achieved mainly through the DM annihilation processes such as
$\psi_X \bar{\psi}_X \rightarrow Z' \rightarrow$ SM particles, $\psi_X \bar{\psi}_X \rightarrow Z' Z'$ and the Higgsstrahlung $\psi_X \bar{\psi}_X \rightarrow Z' h$ depending on the mass spectra.

\begin{figure}[t]
\centering
\includegraphics[width=7.5cm]{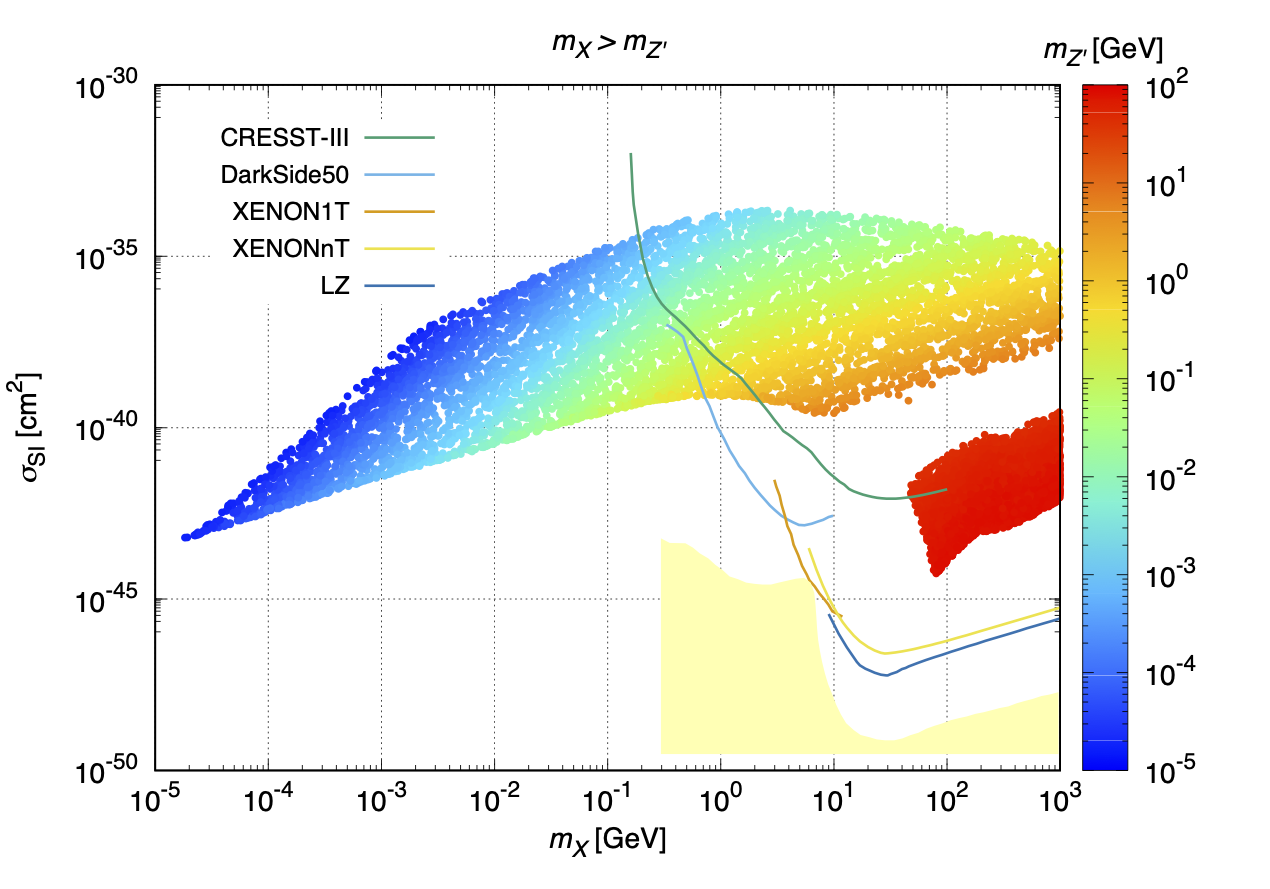}
\includegraphics[width=7.5cm]{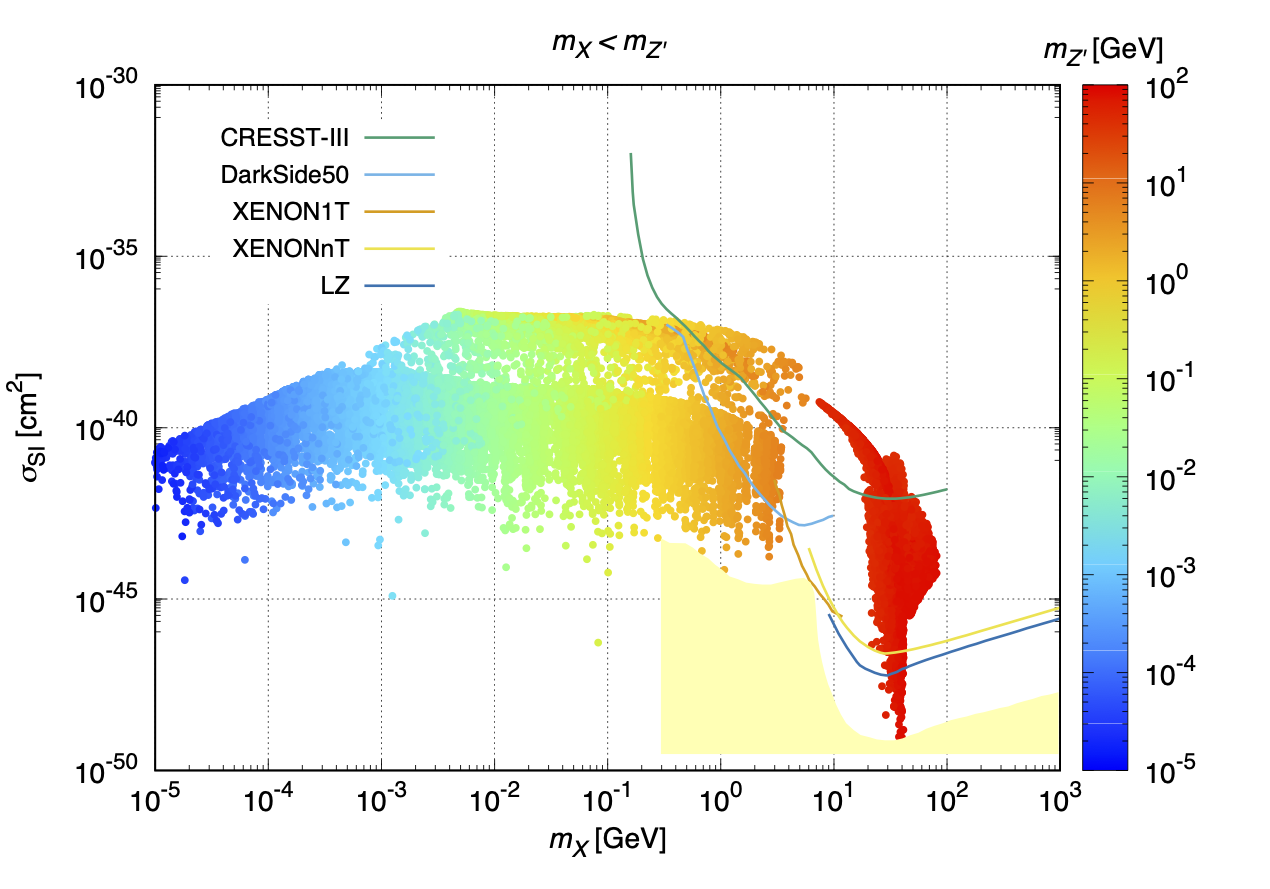}
\caption{Results of the DM-nucleon cross sections for $m_X > m_{Z'}$ ({\it Left}) and  $m_X < m_{Z'}$ ({\it Right}). Discovery Bounds from various direct detection limits are shown together. The yellow shaded region is neutrino background. 
}\label{fig:DD_N}
\end{figure}

\begin{figure}[t]
\centering
\includegraphics[width=7.5cm]{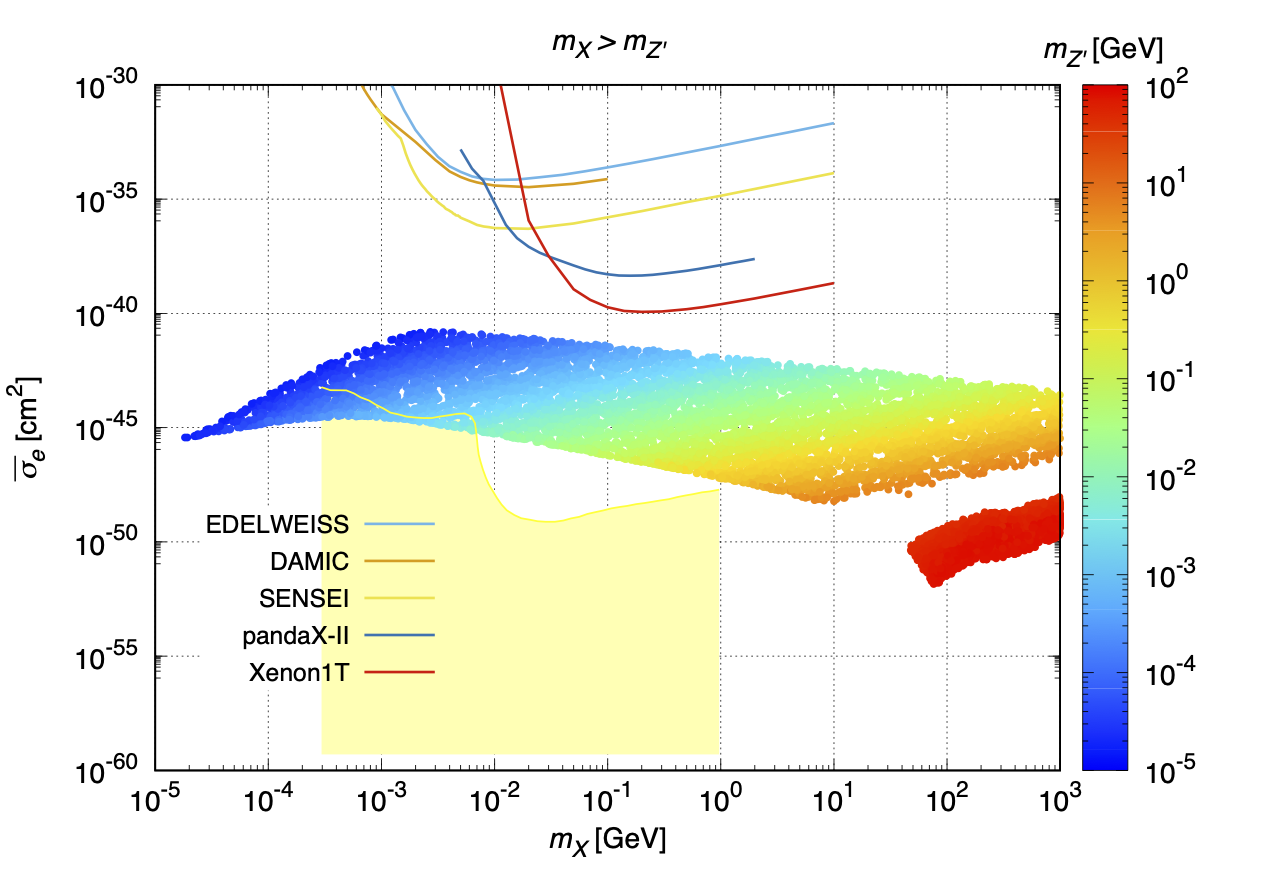}
\includegraphics[width=7.5cm]{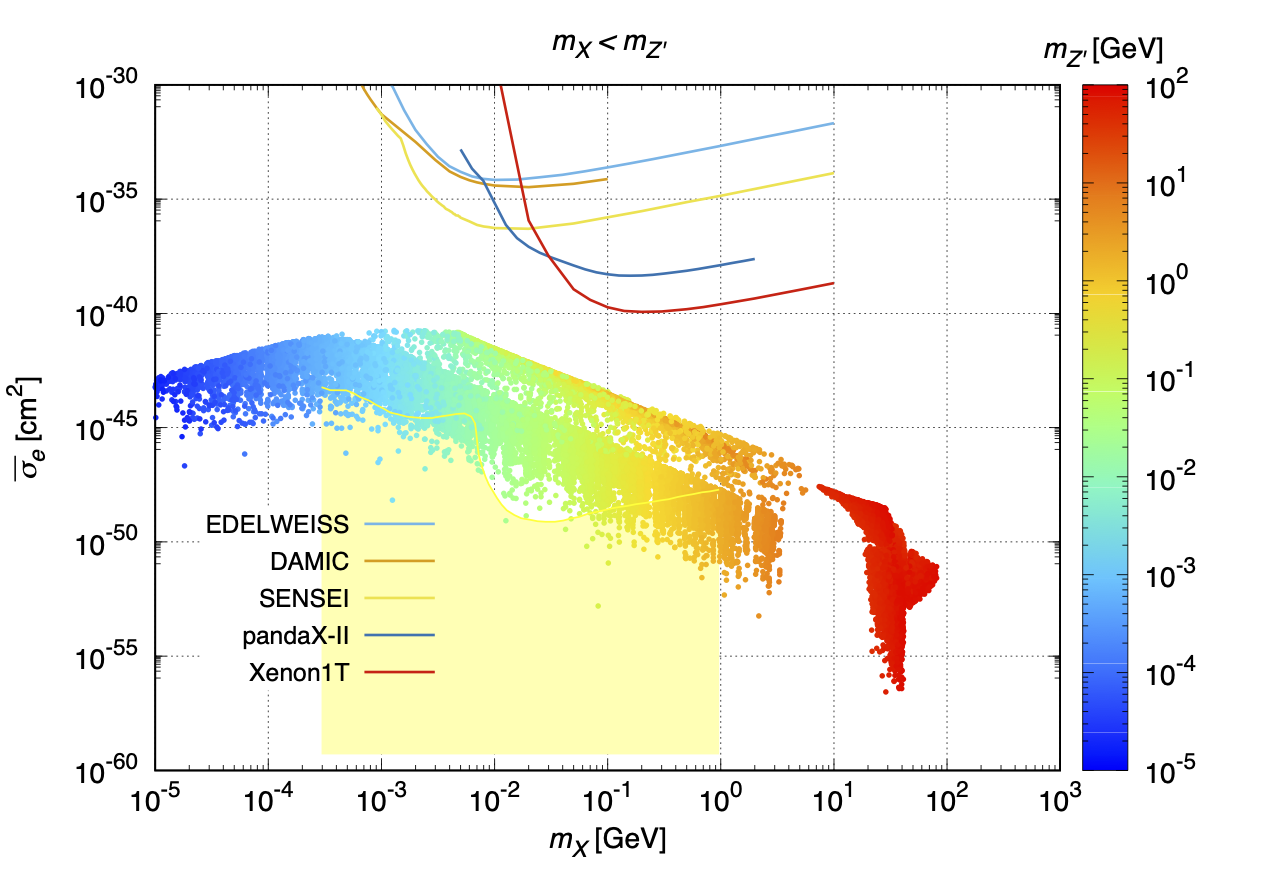}
\caption{Results of the electron-recoil cross sections from  DM for $m_X > m_{Z'}$ ({\it Left}) and  $m_X < m_{Z'}$ ({\it Right}).
Discovery Bounds from various direct detection limits are shown together. The yellow shaded region is neutrino background. 
}\label{fig:DD_e}
\end{figure}

In Fig.~\ref{fig:DD_N}, we show the DM-nucleon cross sections for the parameter sets that are consistent with the observed relic density within 3-$\sigma$ range. We also apply the constraint from the total decay width of the $Z$ boson, which can be modified by the new decay channel $Z\rightarrow \psi_X \bar{\psi}_X$. The parameter sets that give rise to too large DM self interaction are excluded by imposing the `Bullet Cluster' constraint, $\sigma / m_{DM} \lesssim 1 \mathrm{cm^2/g}$ \cite{Randall:2008ppe}. As a result, 
the allowed mass range for the $Z'$ boson is slightly further constrained 
as $m_{Z'} \lesssim 7 \; \mathrm{GeV}$ and $ 43 \;\mathrm{GeV} \lesssim m_{Z'} \lesssim 83 \;\mathrm{GeV}$, compared with the result in the previous section which is analyzed without specifying the DM properties. 

The left figure in Fig.~\ref{fig:DD_N} represents the case that the DM fermion is heavier than $Z'$, $m_X > m_{Z'}$. The experimental bounds from CRESST III \cite{CRESST:2019jnq}, XENON1T \cite{XENON:2020gfr}, XENONnT \cite{XENON:2023sxq}, DarkSide-50 \cite{DarkSide:2022dhx} and LZ \cite{LZ:2022ufs} are shown together. In this case, high mass region for the DM 
fermion is excluded by various direct detection experiments and the DM mass is limited below a few hundred MeV. Accordingly, relatively large $Z'$ mass is prohibited and low mass for $Z'$, less than about a few hundred MeV, is preferred. 

On the other hand, the right figure in Fig.~\ref{fig:DD_N} provides the results for $m_X < m_{Z'}$. In this case, a resonance can be possible in the DM annihilation process $\psi_X \bar{\psi}_X\rightarrow Z' \rightarrow$ SM SM when $2\,m_{X} \sim m_{Z'}$ and the DM pair annihilation rate during the freeze-out is enhanced. This enhancement requires the strength of 
the couplings to be smaller than those for non-resonance cases to fit the observed relic density.  Consequently, the DM-nucleon cross section becomes smaller under the resonance condition and more points survive the limits from the direct detection experiments compared with the previous case of $m_X > m_{Z'}$ . For example, the dip around $m_X\sim 40\, \mathrm{GeV}$ survives the LZ limit for the DM direct detection. In this region, we have `double' resonances such as $2 m_X \sim m_{Z'} \sim m_{Z}$ and the DM-nucleon cross section is suppressed by cancellation between the two diagrams with $Z$ and $Z'$ exchange, respectively.

Since light DM is favored in our model, we should consider electron recoil experiments for DM direct detection. We estimate the cross sections following the Ref.~\cite{Essig:2011nj}, and it turns out that the cross sections survive the limits put by many experiments.
However, the values of the cross sections are small and it is unlikely that the signals from the electron recoil experiments would be detected in the near future. The results together with experimental limits from EDELWEISS \cite{EDELWEISS:2020fxc}, DAMIC \cite{DAMIC:2019dcn}, SENSEI \cite{SENSEI:2020dpa}, PandaX-II \cite{PandaX-II:2021nsg} and XENON1T \cite{XENON:2019gfn} are shown in Fig.~\ref{fig:DD_e}.

\begin{figure}[t]
\centering
\includegraphics[width=12cm]{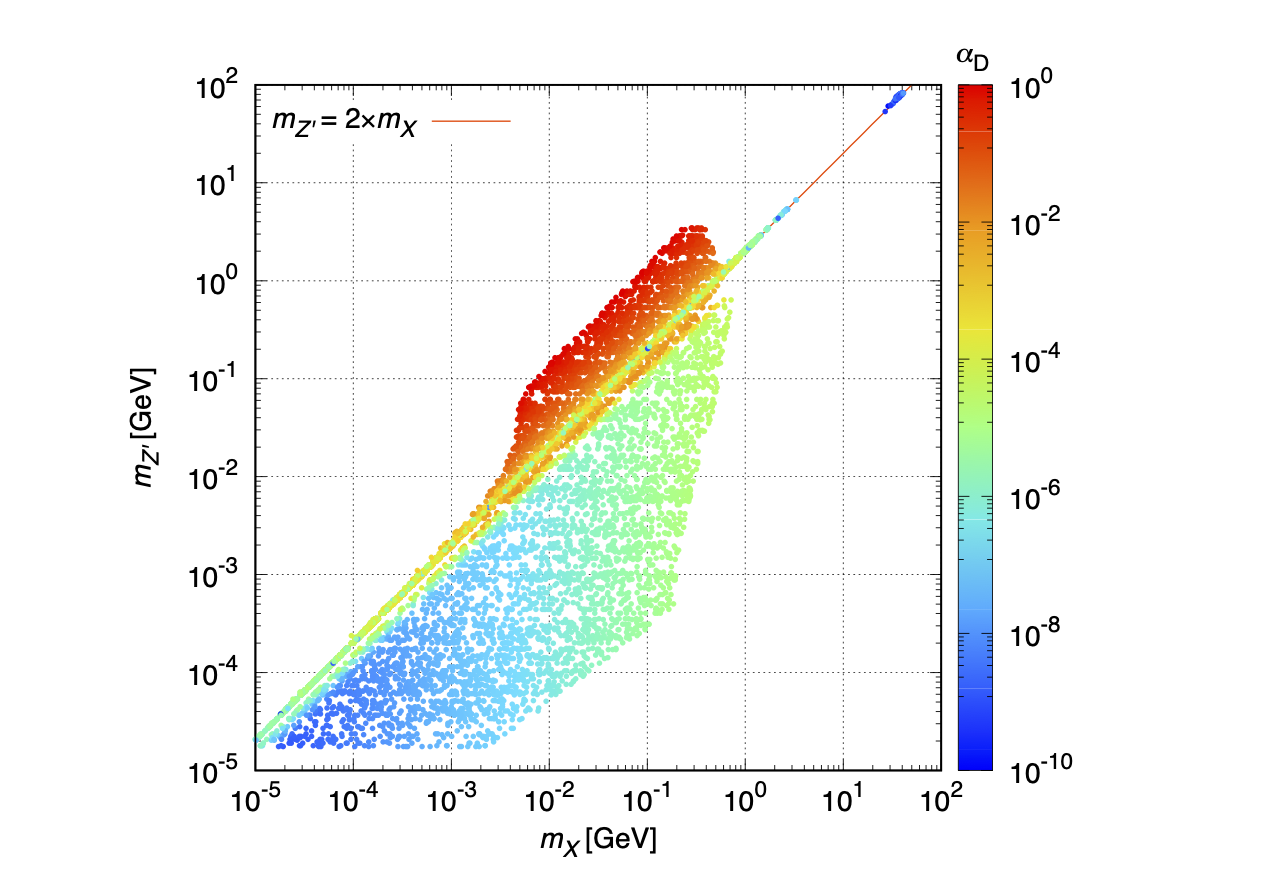}
\caption{
Solution points in $(m_X, m_{Z'})$ plane that survive the discovery limits from the direct detection experiments. The points aligned along the red line represent the resonance points {\it i.e.,} $m_{Z'}=2\,m_X$. The values of dark fine structure constant $\alpha_D\equiv \frac{(g_X X)^2}{4\pi}$ are presented as colour contour.}\label{fig:DM_ZX_sol}
\end{figure}

In Fig.~\ref{fig:DM_ZX_sol}, we show the parameter sets in $\left(m_X,m_{Z'}\right)$ plane that are consistent with the observed values of relic density, the DM self-interaction cross section bound from the Bullet Cluster, total decay width of the $Z$ boson and non-observation of DM in the various direct detection experiments. The DM mass is highly constrained and it would be even below $1$ GeV if we ignore the sharp resonance regions. It is quite interesting that seemingly independent dark sector gives further constraint on the $Z'$ properties. It turns out that the $Z'$ mass of a few tenth of GeV is unfavorable except for the double resonance region 
of $2 m_X \sim m_{Z'} \sim m_{Z}$ while the relatively light $Z'$ boson below a few MeV is preferred. In consequence, the light $Z'$ and DM is quite natural in our model and such properties can be tested in the future collider experiments.

A few comments are in order. Firstly, it is known that for the DM particle as light as  
below 1 MeV, conventional freeze-out mechanism fails \cite{Boehm:2013jpa,Green:2017ybv,Sabti:2019mhn}. However, alternative cosmological scenarios are suggested that can alleviate the problems of sub-MeV DM. See for example Refs.~\cite{Berlin:2017ftj,Berlin:2018ztp}. As an another 
alternative, freeze-in mechanism can be considered \cite{fzin}. Secondly, the production of energetic particles due to self-annihilation of the DM particles in high DM density regions like galactic center can give a serious constraint on the model. We evaluate the velocity-weighted annihilation cross sections $\langle \sigma v \rangle$ especially to $\tau^-\tau^+$ and $b\bar{b}$ since they give the most stringent constraints. We find that all the parameter sets in Fig.~\ref{fig:DM_ZX_sol} give $\langle \sigma v\rangle \sim 10^{-33} - 10^{-26} ~\mathrm{cm^3s^{-1}}$ for both $\tau^-\tau^+$ and $b\bar{b}$ channels and most 
of them stay well below the constraint given by Fermi-LAT which is $\langle \sigma v\rangle_{bb,~\tau\tau} \sim 10^{-26}~ {\rm or}~10^{-27}~\mathrm{cm^3s^{-1}}$ for small DM masses around $2-5$ GeV \cite{Fermi-LAT:2015att} except for few points that are severely fine-tuned through resonances.  Lastly, sub-GeV dark sectors can be constrained from the supernova cooling. We follow the recent comprehensive analysis of Ref.~\cite{Sung:2021swd}, where authors draw the conclusion that even very tiny
self-interaction of the DM particles can trap the DM particles inside the supernova and as a result evade the constraints 
from the energy luminosity of dark emission. We evaluate $\alpha_D \equiv \frac{(g_X X)^2}{4\pi}$ and $\epsilon\sim\sin\theta_X$ up to ${\cal O}(1)$ factor for the DM mass range $1 \mathrm{MeV} \sim 1 \mathrm{GeV}$, and compare 
the results especially with Fig.~4 in Ref.~\cite{Sung:2021swd}. We find that our solution points largely stay outside the exclusion region, which means that DM 
trap mechanism works well and the points in Fig.~\ref{fig:DM_ZX_sol} can circumvent the constraints from the supernova.

\section{Concluding Remarks \label{conclusion}}

In this work, we studied the hidden sector model connected to the SM
through the dark $Z$ boson.
Since the additional SU(2)$_L$ scalar doublet 
with hidden U(1)$_X$ charge mediates 
between the hidden sector and the SM sector,
the EWSB also breaks the hidden sector U(1) symmetry
and generates the massive dark $Z$ boson.
Due to the hidden U(1)$_X$ charge
the additional scalar doublet do not couple to the SM fermions
and the Higgs sector structures of our model 
and the 2HDM of type I are alike.
The CP-odd scalar is eaten up by the massive dark $Z$ boson 
and absent in the Higgs contents of the model.
Instead there are new couplings of the Higgs boson to the dark $Z$ boson.
We explores the Higgs phenomenology of this model
using the public programs {\tt HiggsBounds} and {\tt HiggsSignals}.
The present experimental limits to search the NP in the Higgs sector
strongly constrain the  model  parameter space
together with the electroweak constraints.

We find that our parameter set can satisfy the relic abundance
and the direct detection limits of the DM-nucleon cross section
assuming that the hidden sector fermion is a DM candidate.
Part of the parameter region are more excluded 
by the present direct detection experiments.
The direct detection by the DM-electron scattering is also considered
and we see that the present measurements do not affect our model.
Finally the bullet cluster, the DM annihilation at the galactic center 
and supernova cooling constraints on the DM self-interaction
are discussed to conclude that our model is safe on the whole.

\acknowledgments
This work is supported 
by Basic Science Research Program
through the National Research Foundation of Korea (NRF)
funded by the Ministry of Science, ICT, and Future Planning 
under the Grant No. NRF-2021R1A2C2011003 (DWJ, KYL, CY) and No. NRF-2020R1I1A1A01073770 (CY).
The work of DWJ is also supported in part by NRF-2019R1A2C1089334, NRF-2021R1A2B5B02087078 and the Yonsei University Research Fund of 2022.

\def\PRDD #1 #2 #3 {Phys. Rev. D {\bf#1},\ #2 (#3)}
\def\PRD #1 #2 #3 #4 {Phys. Rev. D {\bf#1},\ No. #2, #3 (#4)}
\def\PRLL #1 #2 #3 {Phys. Rev. Lett. {\bf#1},\ #2 (#3)}
\def\PRL #1 #2 #3 #4 {Phys. Rev. Lett. {\bf#1},\ No. #2, #3 (#4)}
\def\PLB #1 #2 #3 {Phys. Lett. B {\bf#1},\ #2 (#3)}
\def\NPB #1 #2 #3 {Nucl. Phys. B {\bf #1},\ #2 (#3)}
\def\ZPC #1 #2 #3 {Z. Phys. C {\bf#1},\ #2 (#3)}
\def\EPJ #1 #2 #3 {Euro. Phys. J. C {\bf#1},\ #2 (#3)}
\def\JPG #1 #2 #3 {J. Phys. G: Nucl. Part. Phys. {\bf#1},\ #2 (#3)}
\def\JHEP #1 #2 #3 {JHEP {\bf#1},\ #2 (#3)}
\def\JCAP #1 #2 #3 {JCAP {\bf#1},\ #2 (#3)}
\def\IJMP #1 #2 #3 {Int. J. Mod. Phys. A {\bf#1},\ #2 (#3)}
\def\MPL #1 #2 #3 {Mod. Phys. Lett. A {\bf#1},\ #2 (#3)}
\def\PTP #1 #2 #3 {Prog. Theor. Phys. {\bf#1},\ #2 (#3)}
\def\PR #1 #2 #3 {Phys. Rep. {\bf#1},\ #2 (#3)}
\def\RMP #1 #2 #3 {Rev. Mod. Phys. {\bf#1},\ #2 (#3)}
\def\PRold #1 #2 #3 {Phys. Rev. {\bf#1},\ #2 (#3)}
\def\IBID #1 #2 #3 {{\it ibid.} {\bf#1},\ #2 (#3)}


\begin{thebibliography}{99}

\bibitem{Higgs-atlas}
G.~Aad \textit{et al.} [ATLAS],
Phys. Lett. B \textbf{716}, 1-29 (2012)
[arXiv:1207.7214 [hep-ex]].

\bibitem{Higgs-cms}
S.~Chatrchyan \textit{et al.} [CMS],
Phys. Lett. B \textbf{716}, 30-61 (2012)
[arXiv:1207.7235 [hep-ex]].

\bibitem{Bertone:2004pz}
G.~Bertone, D.~Hooper and J.~Silk,
Phys. Rept. \textbf{405}, 279-390 (2005)
[arXiv:hep-ph/0404175 [hep-ph]].

\bibitem{Jung:2020ukk}
D.~W.~Jung, S.~H.~Nam, C.~Yu, Y.~G.~Kim and K.~Y.~Lee,
Eur. Phys. J. C \textbf{80}, no.6, 513 (2020)
[arXiv:2002.10075 [hep-ph]].

\bibitem{Jung:2021bjt}
D.~W.~Jung, K.~Y.~Lee and C.~Yu,
Phys. Rev. D \textbf{105}, no.9, 095023 (2022)
[arXiv:2111.10949 [hep-ph]].

\bibitem{Ko:2013zsa}
P.~Ko, Y.~Omura and C.~Yu,
JHEP \textbf{01}, 016 (2014)
[arXiv:1309.7156 [hep-ph]].


\bibitem{Bechtle:2020pkv}
P.~Bechtle, D.~Dercks, S.~Heinemeyer, T.~Klingl, T.~Stefaniak, G.~Weiglein and J.~Wittbrodt,
Eur. Phys. J. C \textbf{80}, no.12, 1211 (2020)
[arXiv:2006.06007 [hep-ph]];
P.~Bechtle, O.~Brein, S.~Heinemeyer, O.~St\r{a}l, T.~Stefaniak, G.~Weiglein and K.~E.~Williams,
Eur. Phys. J. C \textbf{74}, no.3, 2693 (2014)
[arXiv:1311.0055 [hep-ph]].

\bibitem{Bechtle:2020uwn}
P.~Bechtle, S.~Heinemeyer, T.~Klingl, T.~Stefaniak, G.~Weiglein and J.~Wittbrodt,
Eur. Phys. J. C \textbf{81}, no.2, 145 (2021)
[arXiv:2012.09197 [hep-ph]];
P.~Bechtle, S.~Heinemeyer, O.~St\r{a}l, T.~Stefaniak and G.~Weiglein,
Eur. Phys. J. C \textbf{74}, no.2, 2711 (2014)
[arXiv:1305.1933 [hep-ph]].

\bibitem{Lee:1977eg}
B.~W.~Lee, C.~Quigg and H.~B.~Thacker,
Phys. Rev. D \textbf{16}, 1519 (1977)

\bibitem{Muhlleitner:2016mzt}
M.~Muhlleitner, M.~O.~P.~Sampaio, R.~Santos and J.~Wittbrodt,
JHEP \textbf{03}, 094 (2017)
[arXiv:1612.01309 [hep-ph]].

\bibitem{Grimus:2007if}
W.~Grimus, L.~Lavoura, O.~M.~Ogreid and P.~Osland,
J. Phys. G \textbf{35}, 075001 (2008)
[arXiv:0711.4022 [hep-ph]].

\bibitem{PDG}
R.~L.~Workman \textit{et al.} [Particle Data Group],
PTEP \textbf{2022}, 083C01 (2022)
doi:10.1093/ptep/ptac097

\bibitem{APVSM1}
W. J. Marciano and A. Sirlin, \PRDD 27 552 1983 ; 
{\bf 29}, 75 (1984); {\bf 31}, 213(E) (1985).

\bibitem{APVSM2}
W. J. Marciano and J. L. Rosner, \PRLL 65 2963 1990 ; 
{\bf 68}, 898(E) (1992).

\bibitem{APV}
S. G. Porsev, K. Beloy and A. Derevianko, \PRDD 82 036008 2010 ; 
\PRLL 102 181601 2009 .

\bibitem{ALEPH:2005ab}
S.~Schael \textit{et al.} [ALEPH, DELPHI, L3, OPAL, SLD,
LEP Electroweak Working Group, SLD Electroweak Group
and SLD Heavy Flavour Group],
Phys. Rept. \textbf{427}, 257-454 (2006)
[arXiv:hep-ex/0509008 [hep-ex]].




\bibitem{micromegas} 
G. Blanger, F. Boudjema, A. Goudelis, A. Pukhov and B. Zaldivar,
Comp. Phys. Comm. {\bf 231}, 173 (2018).

\bibitem{Planck:2018vyg}
N.~Aghanim \textit{et al.} [Planck],
Astron. Astrophys. \textbf{641}, A6 (2020)
[erratum: Astron. Astrophys. \textbf{652}, C4 (2021)]
[arXiv:1807.06209 [astro-ph.CO]].

\bibitem{Randall:2008ppe}
S.~W.~Randall, M.~Markevitch, D.~Clowe, A.~H.~Gonzalez and M.~Bradac,
Astrophys. J. \textbf{679}, 1173-1180 (2008)
[arXiv:0704.0261 [astro-ph]].

\bibitem{CRESST:2019jnq}
A.~H.~Abdelhameed \textit{et al.} [CRESST],
Phys. Rev. D \textbf{100}, no.10, 102002 (2019)
[arXiv:1904.00498 [astro-ph.CO]].

\bibitem{XENON:2020gfr}
E.~Aprile \textit{et al.} [XENON],
Phys. Rev. Lett. \textbf{126}, 091301 (2021)
[arXiv:2012.02846 [hep-ex]].

\bibitem{XENON:2023sxq}
E.~Aprile \textit{et al.} [XENON],
[arXiv:2303.14729 [hep-ex]].

\bibitem{DarkSide:2022dhx}
P.~Agnes \textit{et al.} [DarkSide],
Phys. Rev. Lett. \textbf{130}, no.10, 101001 (2023)
[arXiv:2207.11967 [hep-ex]].
\bibitem{LZ:2022ufs}
J.~Aalbers \textit{et al.} [LZ],
[arXiv:2207.03764 [hep-ex]].

\bibitem{Essig:2011nj}
R.~Essig, J.~Mardon and T.~Volansky,
Phys. Rev. D \textbf{85}, 076007 (2012)
[arXiv:1108.5383 [hep-ph]].

\bibitem{EDELWEISS:2020fxc}
Q.~Arnaud \textit{et al.} [EDELWEISS],
Phys. Rev. Lett. \textbf{125}, no.14, 141301 (2020)
[arXiv:2003.01046 [astro-ph.GA]].

\bibitem{DAMIC:2019dcn}
A.~Aguilar-Arevalo \textit{et al.} [DAMIC],
Phys. Rev. Lett. \textbf{123}, no.18, 181802 (2019)
[arXiv:1907.12628 [astro-ph.CO]].

\bibitem{SENSEI:2020dpa}
L.~Barak \textit{et al.} [SENSEI],
Phys. Rev. Lett. \textbf{125}, no.17, 171802 (2020)
[arXiv:2004.11378 [astro-ph.CO]].

\bibitem{PandaX-II:2021nsg}
C.~Cheng \textit{et al.} [PandaX-II],
Phys. Rev. Lett. \textbf{126}, no.21, 211803 (2021)
[arXiv:2101.07479 [hep-ex]].

\bibitem{XENON:2019gfn}
E.~Aprile \textit{et al.} [XENON],
Phys. Rev. Lett. \textbf{123}, no.25, 251801 (2019)
[arXiv:1907.11485 [hep-ex]].

\bibitem{Boehm:2013jpa}
C.~Boehm, M.~J.~Dolan and C.~McCabe,
JCAP \textbf{08}, 041 (2013).

\bibitem{Green:2017ybv}
D.~Green and S.~Rajendran,
JHEP \textbf{10}, 013 (2017).

\bibitem{Sabti:2019mhn}
N.~Sabti, J.~Alvey, M.~Escudero, M.~Fairbairn and D.~Blas,
JCAP \textbf{01}, 004 (2020).

\bibitem{Berlin:2017ftj}
A.~Berlin and N.~Blinov,
Phys. Rev. Lett. \textbf{120}, no.2, 021801 (2018).

\bibitem{Berlin:2018ztp}
A.~Berlin and N.~Blinov,
Phys. Rev. D \textbf{99}, no.9, 095030 (2019).

\bibitem{fzin}
Work in progress.

\bibitem{Fermi-LAT:2015att}
M.~Ackermann \textit{et al.} [Fermi-LAT],
Phys. Rev. Lett. \textbf{115}, no.23, 231301 (2015)
[arXiv:1503.02641 [astro-ph.HE]].

\bibitem{Sung:2021swd}
A.~Sung, G.~Guo and M.~R.~Wu,
Phys. Rev. D \textbf{103}, no.10, 103005 (2021) .



\end{thebibliography}
\end{document}